\documentclass[nofootinbib,twocolumn,showpacs,prd]{revtex4-2}
\usepackage{graphicx}
\usepackage{rotating}
\usepackage{color}
\usepackage{amssymb}

\newcommand{\bmprl}[1]{\mbox{\boldmath $#1$}}
\begin{document}
\title{Comment on ``Scrutinizing \bmprl{\pi\pi} scattering
in light of recent lattice phase shifts''}
\author{
Eef~van~Beveren$^{\; 1\dagger}$
and George~Rupp$^{\; 2}$}
\affiliation{
$^{1}$Centro de F\'{\i}sica da UC,
Departamento de F\'{\i}sica,
Universidade de Coimbra, P-3004-516 Coimbra, Portugal\\
$^{2}$Centro de  F\'{\i}sica e Engenharia de Materiais Avan\c{c}ados,
Instituto Superior T\'{e}cnico, Universidade de Lisboa,
P-1049-001 Lisboa, Portugal
}

\begin{abstract}
In a recent paper by Xiu-Li Gao, Zhi-Hui Guo, Zhiguang Xiao, and
Zhi-Yong Zhou, Phys.\ Rev.\ D {\bf105}, 094002 (2022), here
referred to as I, $S$-wave $\pi\pi$ scattering phase shifts obtained in a
lattice-QCD calculation are analyzed using dispersive $S$-matrix methods. We
question the reliability of the conclusion from this analysis that, for a pion
mass of 391~MeV, the lattice phases favor the presence of both a $\sigma$-meson
bound state and a nearby virtual state. Our main criticism concerns the
neglect of the $S$-wave $K\bar{K}$ channel, which was
considered alongside additional $s\bar{s}$ interpolating fields in the lattice
computation used by the authors of I and also in typical coupled-channel
models. As an illustration, some results from such a recent model are
presented as well. Concluding remarks concern possible
improvements of the analysis in I as well as further model tests.
\end{abstract}

\maketitle
\setcounter{footnote}{1}
\renewcommand{\thefootnote}{\fnsymbol{footnote}}
\footnotetext{Deceased on December 6th, 2022}
We comment on a paper \cite{ARXIV220203124}, hereafter referred to as I,
concerning a two-pole description of $S$-wave ($IJ=00$) pion-pion phase
shifts as determined in the lattice calculation \cite{PRL118p022002} of
the Hadron Spectrum Collaboration (HSC) for a hypothetical pion mass of
391~MeV. The authors of Ref.~\cite{ARXIV220203124} use dispersive
$S$-matrix techniques with crossing-symmetry constraints to model and
simultaneosly fit $\pi\pi$ phases in the channels $IJ=00,20,11$ and for two
different pion masses (391 and 236~MeV) as computed in
Ref.~\cite{PRL118p022002}, henceforth called II, and other papers by the HSC.

Here we focus on the conclusion in I that the $S$-wave $\pi\pi$ phase shifts
mentioned above favor an $S$-matrix description in terms of both a bound-state
(BS) and a relatively nearby virtual-state (VS) pole, instead of only a
BS pole as reported in II. In the following we shall argue why the analysis
of this particular case in I is unreliable. The main reason is the neglect
of the $IJ=00$ $K\bar{K}$ channel, which inevitably couples to the 
$IJ=00$ $\pi\pi$ system and affects the $\sigma$ resonance. In II
both $\pi\pi$ and $K\bar{K}$ two-meson interpolating fields were included,
besides the single-meson operators $u\bar{u}\!+\!d\bar{d}$ and $s\bar{s}$.
Note that the inclusion of $s\bar{s}$ interpolators is crucial to describe
the $f_0(980)$ resonance and the sudden jump of the $S$-wave $\pi\pi$ phases
through 180$^\circ$ in the real situation with the actual pion mass. 
Therefore, the analysis carried out in I is not based on the same
degrees of freedom as in the lattice simulation of II.

So let us first consider the physical $S$-wave $\pi\pi$ phase shifts
and their description in the coupled-channel and fully unitary quark-meson
model of Ref.~\cite{ZPC30p615}. In this paper, the dynamically generated
resonance pole of the $\sigma$ on the second Riemann sheet was actually
accompanied by 4095 other poles, i.e., one on each of the other Riemann
sheets, owing to a total of twelve included meson-meson channels. Such 
a large number of two-meson channels was taken into account \cite{ZPC30p615}
in order to be able to predict $S$-wave $\pi\pi$ phases up to 1.3~GeV, as
well as an additional and also complete scalar-meson nonet in the energy
region 1.3--1.5~GeV. Nevertheless, all these
channels couple to the two bare $^{3\!}P_0$ $u\bar{u}\!+\!d\bar{d}$ and
$s\bar{s}$ channels for the coupled $f_0(500)$-$f_0(980)$ system. Near the
$\sigma$ resonance, the $\pi\pi$ scattering amplitude is well-described
\cite{ZPC30p615} by the dynamically generated pole.  In the lattice
simulation of II, with $m_\pi=391$~MeV, this corresponds to the pole of a
weakly bound state. All the other poles are too far away from the physical
region to be very relevant there. A simple yet
fully unitary toy model in Ref.~\cite{APPB50p1911} qualitatively shows the
contribution of more distant poles to the total amplitude.

\begin{figure}[tb]
\includegraphics[trim = 5mm 10mm 50mm 210mm,clip,width=260pt]
{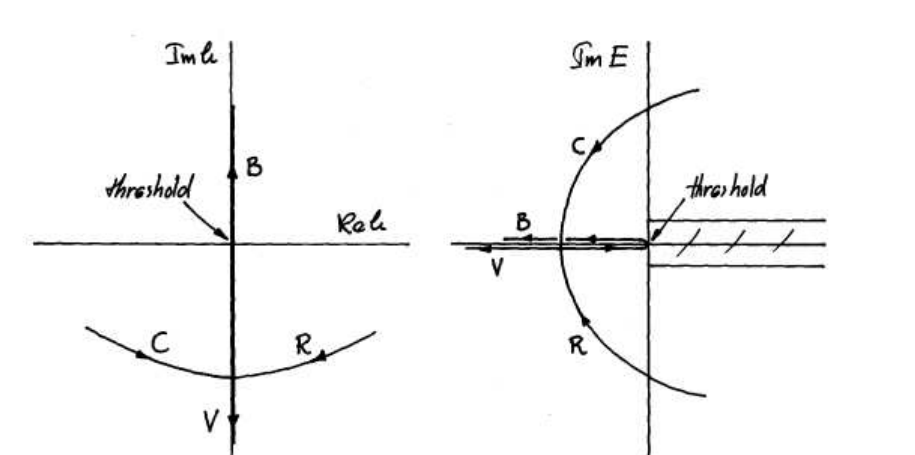}
\caption[]{$S$-matrix pole trajectories in the complex $k$ and $E$
planes, with $E=\sqrt{s}=2\sqrt{k^{2}+m^{2}_{\pi}}$
(reprinted Fig.\ 4.1 of Ref.~\cite{LNP211p331}).}
\label{Fig41}
\end{figure}

Returning to I, FIG.~8 of the paper depicts typical $S$-wave
subthreshold pole trajectories in the complex-energy ($E$) plane.
After both poles hit the real axis, they can either end up as
representing a pair of VSs or one VS and one true BS.
The analysis in I leads to the latter possibility.
However, the existence of a pair of two poles on the real $E$ axis,
viz.\ a VS pole, and either another VS pole or a BS pole,
has been known for decades in a single-channel study;
see e.g.\ Figs.\ 4.1 (reproduced here in FIG.~\ref{Fig41}) and 4.2
of Ref.~\cite{LNP211p331}.
Hence, why the authors stress that
``{\it the description of the $\sigma$ at $m_{\pi}=391$~MeV
as a pair of bound and virtual poles is a novel finding in our study}''
\/is not clear to us. For $S$-wave $K\pi$ and $K\bar{K}$ scattering we
have shown \cite{HEPPH0207022} in detail the pole movements of the BS
and VS poles as a function of the overall coupling (Figs.~6, 7 of
Ref.~\cite{HEPPH0207022}). Similarly, the scalar $D^{\ast}_{s0}(2317)$
meson below the $DK$ threshold is described in Ref.~\cite{PRL91p012003}
and the axial-vector $c\bar{c}$ state $\chi_{c1}(3872)$
slightly below the $S$-wave $D^0\bar{D}^{\star0}$ threshold
in Ref.~\cite{EPJC71p1762}.
Threshold-mass variations were studied in Ref.~\cite{PRD74p037501}.
\begin{figure*}[tb]
\begin{center}
\includegraphics[trim = 0mm 0mm 0mm 0mm,clip,width=17cm,angle=0]
{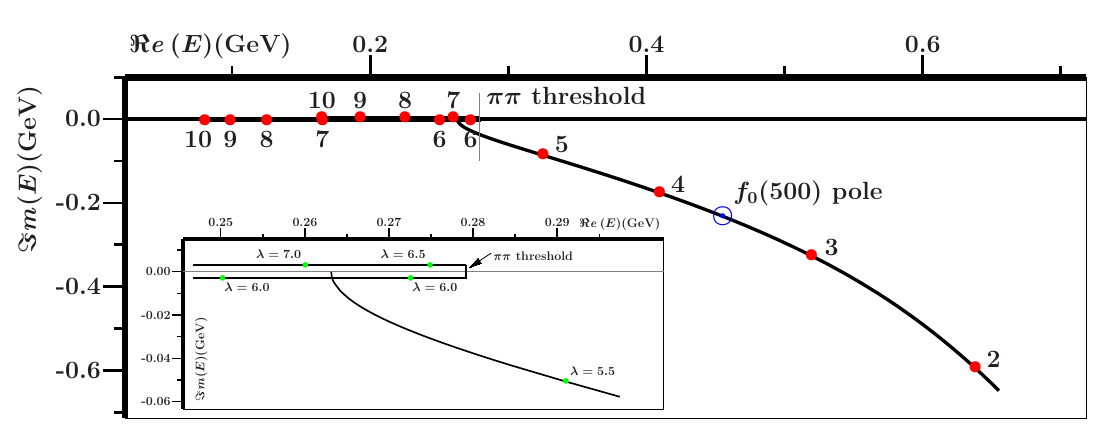} 
\end{center}
\caption{$f_0(500)$ pole trajectory as a function of $\lambda$. The open
circle corresponds to the fitted $\lambda$ value. The inset
shows details of (virtual) bound states, for clarity depicted
slightly (below) above the real axis. Figure is reprinted from Fig.~2
of Ref.~\cite{ARXIV201204994}.}
\label{sigma}
\end{figure*}

In our recent modeling \cite{ARXIV201204994} of $f_0$ resonances
for the physical $m_\pi=139.57$~MeV and with the seventeen $S$- and
$D$-wave meson-meson channels
$\pi\pi$, $K\bar{K}$, $\eta\eta$,
$\eta\eta^\prime$, $\eta^\prime\eta^\prime$, $\rho\rho$, $\omega\omega$,
$K^\star\bar{K}^\star$, $\phi\phi$, $f_0(500)f_0(500)$, $f_0(980)f_0(980)$,
$K_0^\star(700)\bar{K}_0^\star(700)$, and $a_0(980)a_0(980)$, coupled to
the two bare $P$-wave $u\bar{u}+d\bar{d}$ and $s\bar{s}$ channels as
in Ref.~\cite{ZPC30p615}, we found a dynamically generated resonance pole
at $(455-i232)$~MeV on the second Riemann sheet and no VS. This resulted
from a fit to experimental $S$-wave $\pi\pi$ phase shifts up to 1.6~GeV.
The corresponding $f_0(500)$ pole trajectories as a function of the overall
model coupling $\lambda$ are shown (also see Ref.~\cite{ARXIV201204994}) in
Fig.~\ref{sigma}, with the physical $f_0(500)$ pole marked with an open
circle, for $\lambda\!\approx\!3.56$. The figure inset
shows in detail how an $S$-wave resonance pole moves below
the lowest threshold and splits into a pair of VS poles when hitting
the real axis, in agreement with I and Ref.~\cite{LNP211p331}.
Taking a much larger overall coupling, somewhere between 6.0 and 6.5, we
obtain a BS and a VS pole at 0.16 MeV and 62 MeV below the $\pi\pi$ threshold,
respectively. However, for $m_{\pi}=391$~MeV and with the fitted value of
the overall coupling, we find a bound state at 760~MeV, compatible with the
lattice result of 758~MeV in II, but no nearby VS pole. This is to be
contrasted to the dispersive analysis in I for the same $m_\pi\!=\!391$~MeV,
which extracted BS and VS poles at 1~MeV and 73~MeV below threshold,
respectively.

One might question the trustworthiness of our model predictions in
Refs.~\cite{ZPC30p615,ARXIV201204994}, owing to the lack of imposed
crossing-symmetry constraints, despite the remarkably good predictions
for the $\sigma$ pole in both cases. An explanation may be provided
by duality, as remarked in Ref.~\cite{PRL77p2333} (also see 
the pioneering articles in Ref.~\cite{PRL22p562-689}):
\begin{quote}
``\ldots the well-known dual model result for $q\bar{q}$
resonances, that a sum of $s$-channel resonances also describes $t$-
and $u$-channel phenomena.''
\end{quote}
Note that the model calculation referred to in Ref.~\cite{PRL77p2333}
only includes one (bare) $s$-channel state in a unitarized approach, whereas
the unitarized models in Refs.~\cite{ZPC30p615,ARXIV201204994} contain an
infinite tower of such states.

To conclude, we do not question the technical rigor of the
dispersive analysis in I. However, we hope to have made it clear that a
reliable quantitative extraction of possible BS and VS poles
from scattering data, be they experimental or resulting from lattice
simulations, require the consideration of all nearby resonances and inelastic
two-meson channels. In the particular case of $S$-wave $\pi\pi$ phase shifts,
inclusion of the $f_0(980)$ resonance, which strongly affects
\cite{PLB641p265} the phases around 1~GeV,
as well as the $K\bar{K}$ threshold at about 990~MeV is
indispensable. Taking a pion mass of 391~MeV and so generating the $\sigma$
as a weakly bound state does not mean that the influence of the $K\bar{K}$
channel is negligible. Perhaps even more importantly and as
already mentioned above, at the quark level the lattice calculation in II
also included $s\bar{s}$ interpolators besides $u\bar{u}\!+\!d\bar{d}$,
inevitably influencing the resulting $\pi\pi$ phases through the employed
coupled-channel analysis. Moreover, our general experience with VS poles in
multichannel models is that they are much more sensitive to small changes
than BS poles. Finally, the $IJ=00$ $\pi\pi$ lattice phases of II have sizable
statistical error bars, so that any quantitative conclusion from a fit to 
those and other lattice data would already require a lot of caution.

The very lattice results of II appear to confirm the single-pole
scenario, by having extracted a BS (and no nearby VS) very close to our
result for $m_\pi=391$~MeV in the multichannel model of
Ref.~\cite{ARXIV201204994}, with the same $u\bar{u}\!+\!d\bar{d}$,
$s\bar{s}$, $\pi\pi$, and $K\bar{K}$ degrees of freedom.
We do not know whether a coupled-channel generalization of the
dispersive methods in I so as to include besides $\pi\pi$ also the $K\bar{K}$
channel is feasible, but it would certainly
be a topic of interest. For instance, in Ref.~\cite{PRD102p054029}
a three-channel $S$-matrix parametrization with imposed crossing-symmetry
constraints was used to analyze $P$-wave $\pi\pi$ scattering data and
determine excited vector $\rho$ resonances. Furthermore, we plan to do a
comparative study in a simplified version of the model employed in
Ref.~\cite{ARXIV201204994}, which would even allow to explore the behavior of
bound-state and virtual $\sigma$ poles as a continuous function of the pion
mass.

\newcommand{\pubprt}[4]{#1 {\bf #2}, #3 (#4)}
\newcommand{\ertbid}[4]{[Erratum-ibid.~#1 {\bf #2}, #3 (#4)]}
\def\APPB{{Acta Phys.\ Polon.\ B}}
\def\EPJC{{Eur.\ Phys.\ J.\ C}} 
\def\LNP{{Lect.\ Notes Phys.}}
\def\PRD{{Phys.\ Rev.\ D}}
\def\PRL{{Phys.\ Rev.\ Lett.}}
\def\PLB{{Phys.\ Lett.\ B}}
\def\ZPC{{Z.\ Phys.\ C}}

\end{document}